\documentclass{article} 
\usepackage{iclr2021/iclr2021_conference,times}
\usepackage{mathtools}
\usepackage{enumerate}

\usepackage{caption}
\usepackage{amsthm}
\usepackage{subcaption}
\usepackage{bm}
\usepackage{bbold}

\usepackage{subcaption}
\usepackage{graphicx}

\usepackage{amsfonts,amsmath,amssymb}
\usepackage{bbm}
\usepackage{xspace}

\usepackage{multicol}
\usepackage{enumitem}

\usepackage{algorithm}
\usepackage{algpseudocode}

\providecommand{\mc}[1]{\mathcal{#1}}

\providecommand{\mbb}[1]{\mathbb{#1}}

\newcommand{\argmin}{\operatornamewithlimits{argmin}}

\usepackage{subcaption}

\usepackage{hyperref}
\usepackage{url}
\usepackage{cleveref}

\title{Approximately optimal domain adaptation     \\ with Fisher's Linear Discriminant}

\author
{Hayden~S.~Helm$^{*}$ \& Weiwei~Yang \\
Microsoft Research \\
\texttt{hayden@helivan.io} \& \texttt{weiwya@microsoft.com}
\AND
Ashwin De Silva, Joshua T. Vogelstein \& Carey E. Priebe \\
Johns Hopkins University \\
\texttt{\{ldesilv2, jovo, cep\}@jhu.edu}
}

\newcommand\blfootnote[1]{%
  \begingroup
  \renewcommand\thefootnote{}\footnote{#1}%
  \addtocounter{footnote}{-1}%
  \endgroup
}

\iclrfinalcopy 
\begin{document}

\clearpage

\maketitle

\blfootnote{\\
$ ^{*} $ corresponding author}

\vspace{-0.5cm}
\begin{abstract}
    We propose a class of models based on Fisher's Linear Discriminant to address the problem of learning a classifier from a collection of source tasks for a resource-constrained target task. 
    The class is the convex combinations of an average of the source task classifiers and a classifier trained on the limited data available for the target task.
    We derive the expected loss of an element in the class with respect to the target distribution for a specific generative model and subsequently propose a computable approximation of the loss.
    We demonstrate in both analytical and real-data experiments that the element of the proposed class that minimizes the approximated risk is able to exploit a natural bias-variance trade-off in task space. 
    We conclude by discussing further applications, limitations, and potential future directions.
\end{abstract}
\section{Introduction}

For problems with limited task-specific data, supervised machine learning models often fail to generalize well. 
Classically, practitioners operating in these settings will choose a model that is appropriately expressive given the amount of data available. That is, they use a model that effectively exploits the ``bias-variance" trade-off \citep{vonluxburg2008statistical}.
Modern machine learning approaches such as transfer learning~\citep{pan2010survey, weiss2016survey}, domain adaptation~\citep{ sun2015survey}, meta-learning~\citep{vilalta2002perspective,  vanschoren2019meta,finn2019online, hospedales2021meta}, and continual learning~\citep{van2019three, hadsell2020embracing,de2021continual,  vogelstein2022representation} attempt to mitigate the lack of task-specific data by leveraging information from a collection of available source tasks. These approaches are ineffective when the task of interest is sufficiently different from the source tasks. 

In this paper we study a data-adaptive method that can interpolate between the classical and modern approaches for a specific set of classifiers: when the amount of available task-specific data is large and the available source tasks are sufficiently different then the method is equivalent to the classical single task approach; conversely, when the amount of available task-specific data is small and the available source tasks are similar to the task of interest then the method is equivalent to the modern approach. 

At a high level, our proposed method is designed in the context of a set of classifiers based on Fisher's Linear Discriminant (``FLD")~\citep{izenman2013linear, devroye2013probabilistic}. 
Each element in the class is a convex combination of i) an average of linear classifiers trained on source tasks and ii) a classifier trained only on data from a new target task.
Given the set of classifiers, we derive the expected risk (under 0-1 loss) of an element in the class under particular generative assumptions, approximate the risk using the appropriate limit theory, and select the classifier that minimizes this approximated expected risk.
By approximating the expected risk, we are able to simultaneously take advantage of the relationship between the source tasks and the target task and the new information available related to the target task.

We focus on FLD, as opposed to more complicated classification techniques,
due to its popularity in low resource settings. 
For example, our motivating setting is the physiological prediction problem -- broadly defined as any setting that uses biometric or physiological data (e.g., EEG, ECG, breathing rate, etc.) or any derivative thereof to make predictions related to the state of a person -- where polynomial classifiers and regressors with expert-crafted features are still the preferred performance baselines \citep{eeg-da}.

The rest of the paper is organized as follows: We first review relevant aspects of the domain adaptation, physiological prediction, and task similarity literature. 
We then describe our problem setting formally, introduce notation, and motivate our method by reviewing the distributional assumptions for which FLD is optimal under 0-1 classification loss.
We subsequently make the relationship between the source distributions and target distribution explicit by leveraging the sufficiency of the FLD projection vector.
We define the set of classifiers based on this relationship, derive an expression for the expected risk of a general element in this set, and propose a computable approximation to it that can be used to find the optimal classifier in the set.
Finally, we study the effect of different hyperparameters of the data generation process on the performance of the approximated optimal classifier relative to model i) and model ii) before applying it to three physiological prediction settings.

\subsection{Related works}

\subsubsection{Connection  to Domain Adaptation Theory}

The problem we address in this work can be framed as a domain adaptation problem with multiple sources. While a rich body of literature~\citep{ben2010theory, mansour2008domain,duan2012domain,sun2015survey,guo2018multi, zhang2015multi,zhao2018adversarial,de2023value} has studied this setting, our work shares the most resemblance with the theoretical analysis discussed in~\citep{mansour2008domain}. They study the combination of the source classifiers and derive a hypothesis that achieves a small error with respect to the target task. In their work, they assume the target distribution is a mixture of the source distributions. Our work, on the other hand, combines the average source classifier with the target classifier under the assumption that the classifiers originate from the same distribution on the task level. Indeed, the explicit relationship that we place on the source and target projection classifiers allows us to derive an analytical expression of the target risk that does not rely on the target distribution being a mixture of source~distributions.


\subsubsection{Domain Adaptation for Physiological Prediction Problems}
Domain adaptation and transfer learning are ubiquitous in the physiological prediction literature due to large context variability and small in-context sample sizes. See, for example, a review of EEG-inspired methods~\citep{zhang2020application} and a review of ECG-inspired methods~\citep{ecg-da}. Most similar to our work are methods that combine general-context data and personalized data~\citep{nkurikiyeyezu2019effect}, or weigh individual classifiers or samples from the source task based on similarities to the target distribution~\citep{zadrozny2004proceedings, azab2019weighted}. Our work differs from~\citep{nkurikiyeyezu2019effect}, for example, by explicitly modeling the relationship between the source and target tasks. This allows us to derive an optimal combination of the models as opposed to relying strictly on empirical~measures.

\subsubsection{Measures of Task Similarity}
Capturing the difference between the target task and the source tasks is imperative for data-driven methods that attempt to interpolate between different representations or decision rules. We refer to attempts to capture the differences as measures of task similarity measures. Generally, measures of task similarity can be used to determine how relevant a pre-trained model is for a particular target task~\citep{bao2018information, tran2019transferability, leep, de2023value} or to define a taxonomy of tasks~\citep{taskonomy}.

In our work, the convex coefficient $ \alpha $ parameterizing the proposed class of models can be thought of as a measure of model-based task dissimilarity between the target task and the average-source task---the farther the distribution of the target projection vector is from the distribution of the source projection vector the larger the convex coefficient. 
Popular task similarity measures utilize information theoretic quantities to evaluate the effectiveness of a pre-trained source model for a particular target task such as H-score~\citep{bao2018information}, NCE~\citep{tran2019transferability}, or LEEP~\citep{leep}. This collection of work is mainly empirical and does not place explicit generative relationships on the source and target tasks. 
Other statistically inspired task similarity measures, like ours, rely on the representations induced by the source and target classifiers such as partitions~\citep{tasksim} and other model artifacts \citep{baxter2000model,ben2003exploiting, xue2007multi}. Similar ideas have been used to leverage the presence of multiple tasks for ranking~\citep{helm2021leveraging}.


\subsection{Problem setting}

The classification problem discussed herein is an instance of a more general statistical pattern recognition problem~\citep[Chapter~1]{devroye2013probabilistic}: Given training data  $ \{(X_{i}, Y_{i})\}_{i=1}^{n} \in \left(\mathcal{X} \times \{1, \hdots, K\}\right)^{n} $ assumed to be $i.i.d.$ samples from a classification distribution $ \mathcal{P} $, construct a function $ h_{n} $ that takes as input an element of $ \mc{X} $ and outputs an element of $ \{1, \hdots, K\} $ such that the expected loss of $ h_n $ with respect to $ \mc{P} $ is small. With a sufficient amount of data and suitably defined loss, there exists 
a classifier $ h_{n} $ that has statistically minimal expected loss for any given $ \mc{P} $. In the prediction problems like the physiological prediction problem, however, there is often \textit{not} enough data from the target task to adequately train classifiers and we assume, instead, that there is auxillary data (or derivatives thereof) from different contexts available that can be used to improve the expected loss \citep{pan2010survey}.

In particular, given $ \{(X_{i}^{(j)}, Y_{i}^{(j)})\}_{i=1}^{n} $ assumed to be $ i.i.d. $ samples from the classification distribution $ \mc{P}^{(j)} $ for $ j \in \{0, \hdots, J\} $, we want to construct a classifier $ h^{(0)} $ that minimizes the expected loss with respect to the target distribution $ \mc{P}^{(0)} $. We refer to the classification distribution $ \mc{P}^{(j)} $ as a source distribution for $ j \in \{1, \hdots, J\} $. Note that for other modern machine learning settings the classifier $ h^{(0)} $ is constructed to optimize joint loss functions with respect to $ \mc{P}^{(0)}, \hdots, \mc{P}^{(J)} $\citep{geisa2022theory}.

Generally, for the classifier $ h^{(0)} $ to improve upon the task-specific classifier $ h_{n} $, the source distributions need to be related to the target distribution such that the information learned when constructing the mappings from the input space to the label space in the context of the source distributions can be ``transferred" or ``adapted" to the context of the target distribution~\citep{baxter2000model}.

\section{Method}
\label{sec:generative-model}

Our goal is to develop a
classifier that can leverage information from data from both the target and source distributions. 
For this purpose, we first make distributional assumptions on the data from a single task and then explicitly describe the assumed relationship between the target and source tasks.

\subsection{Motivating distributional assumptions}

In particular, we assume that $ \mc{P}^{(j)} $ is a binary classification distribution that can be described as follows:
\begin{equation}
\label{eq:model-assumptions}
    \mc{P}^{(j)} = \pi^{(j)} \;\mathcal{N}\left(\nu^{(j)}, \Sigma^{(j)}\right) + (1 - \pi^{(j)}) \;\mathcal{N}\left((-1)\nu^{(j)}, \Sigma^{(j)}\right); \quad \text{for } j \in \{0, \hdots, J\}.
\end{equation} To be explicit, $ \mc{P}^{(j)} $ is a mixture of two Gaussians such that the midpoint of the class conditional means is the origin and that the class conditional covariance structures are equivalent. Note that $\mc{P}^{(j)}$ is uniquely parameterized by $\nu^{(j)}, \Sigma^{(j)},$ and $\pi^{(j)}$ and that the shared conditional covariance is a standard assumption when using linear models. 

Let $ \mathbbm{1}\{s\} $ be the indicator function that returns $ 1 $ if $ s $ is true and $ 0 $ otherwise. Recall that under the generative assumptions described above the linear classifier 
\begin{align*}
    h_{FLD}(x) = \mathbbm{1}\left\{\omega^{\top} x > c\right\},
\end{align*} where
\begin{align}
\label{eq:fld-parameters}
    \omega = (\Sigma_0 + \Sigma_1)^{-1} (\nu_1 - \nu_0) \quad \quad \text{and} \quad \quad c = \omega^\top \left(\nu_0 + \nu_1\right) + \log \frac{\pi_0}{\pi_1} 
\end{align} is optimal under 0-1 loss for distributions of the form described in Eq. \eqref{eq:model-assumptions}.

We further restrict our analysis to settings with $ \pi = 0.5 $ where the definitions in Eq. \eqref{eq:fld-parameters} reduce to $ \omega = \frac{1}{2}\Sigma^{-1} (\nu_1 - \nu_0) $ and $ c = 0 $. With this restriction, $ h_{FLD} $ depends only on the projection vector $ \omega $. Since the optimal classifier for a task is parameterized solely by its projection vector, we consider the projection vector as the sole parameter for the task itself.  Thus, to describe a relationship between classification tasks in our setting we need only to describe a relationship on their optimal projection vectors. 

Recall that the von Mises-Fisher (vMF) distribution \citep{fisher1993statistical}, denoted by $\mc{V}(\mu, \kappa)$, has realizations on the $ d $-sphere and is completely characterized by a mean direction vector $ \mu \in \mathbb{R}^{d} $ and a concentration parameter $ \kappa \in \mathbb{R}_{\ge 0} $. When the concentration parameter is close to $ 0 $ the vMF distribution is close to a uniform distribution on the $ d$-sphere. When the concentration parameter is large, the vMF distribution resembles a normal distribution with mean $ \mu $ and a scaled isotropic variance proportional to the inverse of $ \kappa $. 

For our analysis we assume that the optimal projection vectors $ \omega^{(0)},  \omega^{(1)}, \hdots, \omega^{(J)}  \stackrel{iid}{\sim} \mathcal{V}(\mu, \kappa) $ for unspecified $ \mu $ and $ \kappa $. Given the assumed equality of the class conditional covariance structures and that the class conditional means are additive inverses, $ \omega^{(j)} $ being a unit vector forces an additional constraint on the relationship between $ \nu^{(j)} $ and $ \Sigma^{(j)} $ in the context of Eq. \eqref{eq:model-assumptions} -- namely that $ || (\Sigma^{(j)})^{-1}\nu^{(j)} ||_{2} = 1 $. In the simulation settings below, the generative models adhere to this constraint. In practical applications we can use the (little) training data that we have access to force our estimates of $ \nu $ and $ \Sigma $ to be conformant. 

\subsection{A Class of Linear Classifiers}

\begin{figure*}[h]
    \centering
    \includegraphics[width=0.5\linewidth]{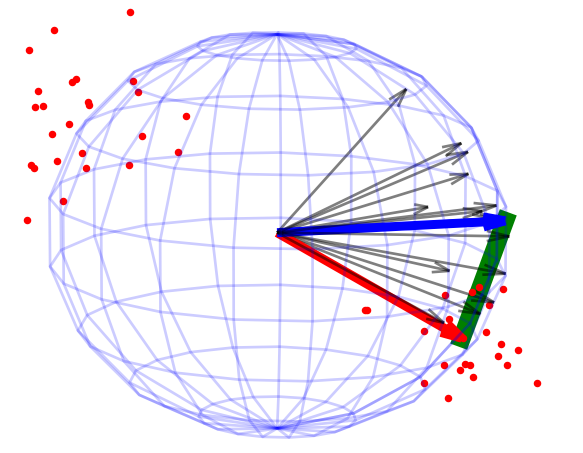}
    \caption{A geometric illustration of the generative assumptions, information constraints, and the model class under study. All vectors are unit vectors. The red dots represent the data from the target distribution, the red arrow represents an estimate of the projection vector for the target distribution, the grey arrows represent source projection vectors, the blue arrow represents the average-source projection vector, and the green line interpolating between the blue and red arrows represents the possible end points of a convex combination of the red and blue arrows.}
    \label{fig:illustrative-figure}
\end{figure*}

With the generative assumptions described above, we define a class of classifiers $ \mc{H} $ that can leverage both the information in the source projection vectors and the target projection vector:
\begin{equation*}
    \mc{H} := \left\{h_{\alpha}(x) = \mathbbm{1} \left\{  \left( \underbrace{\alpha \omega^{(0)} + (1 - \alpha) \sum_{j=1}^{J} \omega^{(j)}}_{\omega_{\alpha}}\right)^{\top}x > 0\right\} : \alpha \in [0,1] \right\}.
\end{equation*} The set $ \mc{H} $ is exactly the classifiers parameterized by the convex combinations of the target projection vector and the sum of the source projection vectors. We refer to this convex combination as $ \omega_{\alpha} $.  Letting $ \bar{\omega} := \frac{1}{J}\sum_{j=1}^{J} \omega^{(j)} $, we note that $ \omega_{\alpha} $ can be reparametrized in the context of the vMF distribution with the observation that
\begin{align}
\label{eq:vmf-perspective}
    (1-\alpha) \sum_{j=1}^{J} \omega^{(j)} = (1-\alpha) \frac{J||\bar{\omega}||}{J||\bar{\omega}||} \sum_{j=1}^{J}\omega^{(j)} = (1-\alpha)J||\bar{\omega}||\frac{\bar{\omega}}{||\bar{\omega}||} 
    = J(1-\alpha)||\bar{\omega}|| \; \hat{\mu},
\end{align} where $ \hat{\mu} = \bar{\omega} / ||\bar{\omega} || $ is the maximum likelihood estimate for the mean direction vector of the vMF distribution. By letting $ \alpha \gets \frac{\alpha}{\alpha + J (1-\alpha)||\bar{\omega}||}$ 
we maintain the same set $ \mc{H} $ but make the individual classifiers more amenable to analysis. Figure \ref{fig:illustrative-figure} illustrates the geometry of $ \mc{H} $ for $ d = 3 $.

With the parameterization implied by the right-most expression of Eq. \eqref{eq:vmf-perspective}, we view different decision rules in $ \mc{H} $ as elements along a classical bias-variance trade off curve \textit{in task space} parameterized by $ \alpha $ \citep{belkin2019reconciling}. In particular, when the amount of data available from the target distribution is small, the projection induced by an $ \alpha $ value closer to 1 can be interpreted as a high variance, low bias estimate of the target projection vector. Conversely, an $ \alpha $ value of 0 can be interpreted as a low variance, high bias estimate. In situations where the concentration parameter $ \kappa $ is relatively large, for example, we expect to prefer combinations that favor the average-source vector. We discuss this in more detail in Section \ref{subsec:simulations}.

\subsection{Approximating optimality}\label{subsec:approximating-optimality}


We define the optimal classifier $ h_{\alpha^{*}} \in \mc{H} $ as the classifier that minimizes the expected risk with respect to the target distribution $ \mc{P}^{(0)} $. Given the projection vectors $ \{\omega^{(j)}\}_{j=1}^{J} $ and the target class conditional mean and covariance, $ \nu^{(0)} $ and $ \Sigma^{(0)} $, the risk (under 0-1 loss) of a classifier $ h_{\alpha} \in \mc{H} $ is
\begin{equation*}
\label{risk}
    R \left( h_{\alpha} \mid \{\omega^{(j)}\}_{j=1}^{J}, \nu^{(0)}, \Sigma^{(0)} \right) = \Phi \left( \frac{-\omega_{\alpha}^\top \nu^{(0)}}{\sqrt{\omega_{\alpha}^\top \Sigma^{(0)} \omega_{\alpha}}} \right)
\end{equation*} 
for $\mathcal{P}^{(0)} $ of the form described in Eq. \eqref{eq:model-assumptions} and where $ \Phi $ is the cumulative distribution function of the standard normal distribution. The derivation is given in Appendix \ref{sec:app-derivation}. In practice, the source projection vectors, target class conditional mean and covariance structure are all estimated. 

We define the expected risk of $h_{\alpha}$ as
\begin{equation}
\label{eq:risk-integral}
    \mathcal{E}(h_{\alpha}) = \mbb{E}_{\omega_{\alpha}}  \left[ R \left(h_{\alpha} \mid \{\omega^{(j)}, \}_{j=1}^{J}, \nu^{(0)}, \Sigma^{(0)}\right)\right].
\end{equation}
Despite the strong distributional assumptions we have in place, the expected risk is still too complicated to analyze entirely.
Instead, we can approximate $ \mathcal{E}(h_{\alpha}) $ by sampling from the distribution of $ \omega_{\alpha} $ (derived in Section \ref{subsec:distribution-of-omega-alpha}) using the plug-in estimates for $ \nu^{(0)} $ and $ \Sigma^{(0)} $.  

The entire procedure for calculating the optimal $ \alpha $ with the approximated risk function is outlined in Algorithm \ref{alg:cap}. For the remainder of this section, we use $ \hat{t} $ to denote an estimate of the parameter $ t $.



\begin{algorithm}[t]
\caption{Calculating the optimal convex coefficient}\label{alg:cap}
\label{alg:optimal}
\begin{algorithmic}[1]
\Require target task class conditional mean $\hat \nu_0^{(0)}$ and $\hat \nu_1^{(0)} $, target task class conditional covariance $\hat \Sigma^{(0)} $, normalized source proj. vectors $\{ \hat \omega^{(j)} \}_{j=1}^J$, grid step size $ h $, the number of bootstrap samples $ B $.
\State
$ \hat \omega^{(0)} \gets \textproc{Normalize} \left( \frac{1}{2} (\hat \Sigma^{(0)})^{-1} (\hat \nu_1^{(0)} - \hat \nu_0^{(0)}) \right)$
\Comment{Estimate the target proj. vector}
\State
$ \hat{\mu} \gets$ \textproc{Normalize}$\left(\frac{1}{J} \sum_{j=1}^{J} \hat \omega^{(j)} \right)$
\Comment{Estimate vMF mean direction vector}
\State
$ \hat \Psi \gets $ \textproc{Approx-Cov}$\left(\{\hat{\mu}^{(j)}\}_{j=1}^{J}\right) $
\Comment{Covariance of $ \hat{\mu} $}(see Eq. \eqref{eq:vmf-cov})
\State
$ \hat \Sigma_{\omega} \gets $ \textproc{Covariance}$\left(\hat \omega^{(0)} \right) $ \Comment{Covariance of the target proj. vector (see \cref{subsec:distribution-of-omega-alpha})}
\For{each $\alpha \in \{0, h, 2h, \hdots, 1-h, 1\}$}
    \State
    $ \hat \omega_{\alpha} \gets \left(\alpha \; \hat \omega^{(0)} + (1-\alpha)\; \hat{\mu} \right)$ \Comment{Average convex combination}
    \State
    $ \hat \Sigma_{\alpha} \gets \alpha^{2} \hat \Sigma_{\omega} + (1-\alpha)^{2}\hat \Psi $
    \Comment{Covariance of average convex combination}
    \For{each $ b $ in $\{1, .. , B\} $}
        \State
        $ \omega_{b} \gets \mc{N}\left(\hat \omega_{\alpha}, \hat \Sigma_{\alpha}\right)$
        \Comment{Sample from appropriate normal distribution}
        \State
        $r_{b} \gets \Phi \left( - \frac{\omega_{b}^{\top} \hat \nu^{(0)}}{\sqrt{\omega_{b}^{\top} \hat \Sigma^{(0)} \; \omega_{b}}} \right) $ \Comment{Calculate error for sample}
    \EndFor
    \State $\hat{\mathcal{E}}\left({\alpha}\right) \gets \frac{1}{B} \sum_{b=1}^{B} r_{b} $ \Comment{Calculate risk}
\EndFor
\State $\alpha^\ast \gets \argmin_{\alpha} \hat{\mathcal{E}} (\alpha)$ \Comment{Select optimal alpha}
\end{algorithmic}
\end{algorithm}

\subsection{Deriving the asymptotic distribution of $ \hat{\omega}_{\alpha} $}
\label{subsec:distribution-of-omega-alpha}

We are interested in deriving a data-driven method for finding the element of $ \mc{H} $ that performs the best on the target task. For this, we rely on the asymptotic distribution of $ \hat{\omega}_{\alpha} = \alpha \hat{\omega}^{(0)} + (1-\alpha) \hat{\mu} $. 

First, we consider the estimated target projection vector $\hat \omega^{(0)} = \frac{1}{2} (\hat \Sigma^{(0)})^{-1} (\hat\nu_1^{(0)} - \hat\nu_0^{(0)})$ as a product of the independent random variables, $A := n (\hat \Sigma^{(0)})^{-1}$ and $\tau := \frac{1}{2}(\hat\nu_1^{(0)} - \hat\nu_0^{(0)})$. We next note that $A \sim W_{d}(n, \Sigma^{(0)})$ is distributed according to a Wishart distribution with $n$ degrees of freedom and scatter matrix $\Sigma^{(0)}$. Further, $\tau \sim \mc{N}_d(\nu^{(0)}, \Sigma^{(0)} / n)$ is normally distributed. 
Thus, for large $ n $ the random vector $nA^{-1}\tau$ has the asymptotic distribution given by
\[ 
\sqrt{n} \left( nA^{-1}\tau - (\Sigma^{(0)})^{-1} \nu \right) \stackrel{d}{\longrightarrow} \mc{N}_d(0, \tilde{\Sigma})
\]
where $\tilde{\Sigma} = (1 + (\nu^{(0)})^\top (\Sigma^{(0)})^{-1} \nu^{(0)}) (\Sigma^{(0)})^{-1} - (\Sigma^{(0)})^{-1} \nu^{(0)} (\nu^{(0)})^\top (\Sigma^{(0)})^{-1} $ \citep{kotsiuba2016asymptotic}. It follows that $\hat \omega^{(0)}$ is aymptotically distributed according to a normal distribution with mean $\omega^{(0)} = (\Sigma^{(0)})^{-1} \nu^{(0)}$ and covariance matrix $\Sigma_{\omega} := \tilde{\Sigma} / n$. 

Next, we observe that $\hat \mu$ is the sample mean direction computed from $J$ i.i.d samples drawn from a $\mathcal{V}(\mu, \kappa)$. For large $J$ we have $\hat \mu$ asymptotically distributed as a normal distribution with mean $\mu$ and covariance $\Psi$ given by 
\begin{equation}
\label{eq:vmf-cov}
\Psi = \left( \frac{1 - \frac{1}{J} \sum_{j=1}^{J} (\mu^\top \omega^{(j)})^2}{J \| \bar\omega \|} \right)^{1/2} I_d,
\end{equation}
where $I_d$ is the $d \times d$ identity matrix \citep{fisher1993statistical}. 

Finally, since $\hat \omega$ and $\hat \mu$ are independent and asymptotically normally distributed, for large $n$ and $J$, we have
\begin{equation*}
\label{risk-integral}
\hat{\omega}_{\alpha} \sim \mathcal{N}\left(\underbrace{\alpha \omega^{(0)} + (1-\alpha) \mu}_{\omega_{\alpha}}, \underbrace{\alpha^2 \Sigma_{\omega} + (1-\alpha)^{2} \Psi}_{\Sigma_{\alpha}}\right).
\end{equation*}

\noindent We use samples from this asymptotic distribution when evaluating the risk function described in Eq. \eqref{eq:risk-integral} and use the $ \alpha $ that minimizes it to choose the classifier in $ \mc{H} $ to deploy. We describe the exact procedure for calculating the optimal classifier $ h_{\alpha^{*}} $ in Algorithm \ref{alg:optimal}, where \texttt{Approx-Cov} returns $ \Psi $ as described in Eq.\eqref{eq:vmf-cov}.
\section{Simulations}
\label{subsec:simulations}

In this section we first validate our method by comparing our approximation to the true-but-analytically-intractable risk to the empirical risk under a fixed set of generative model parameters. We then study the effect of different generative model parameters on the relative risks of the target classifier, the average-source classifier, and the approximately optimal classifier. For each simulation setting we report the expected accuracy (i.e., 1 minus the expected risk) and the optimal convex coefficient $ \alpha^{*} $.

For the purposes of our simulations, we let $ d $ be the dimensionality of the data and $ n $ be the number of samples from the target distribution. Without loss of generality, we consider a von Mises-Fisher distribution with mean direction $\mu = [1, 0_{d-1}]^\top$ and concentration parameter $\kappa$. We fix the mixing coefficient $ \pi^{(j)} = 0.5 $ and the class-conditional covariance $ \Sigma^{(j)} = I_{d} $ for all task distributions for all simulation settings. For each Monte Carlo replicate and for each simulation setting, we sample $ \nu^{(0)} $ and $ \{\omega^{(j)}\}_{j=1}^{J} $ from $ \mathcal{V}(\mu, \kappa) $. Finally, for each simulation setting we report the mean accuracy over 1,000 iterations and, hence, the standard error of each estimate is effectively zero. 

\subsection{Validating the approximation}

\begin{figure*}[h]
    \centering
    \includegraphics[width=\linewidth]{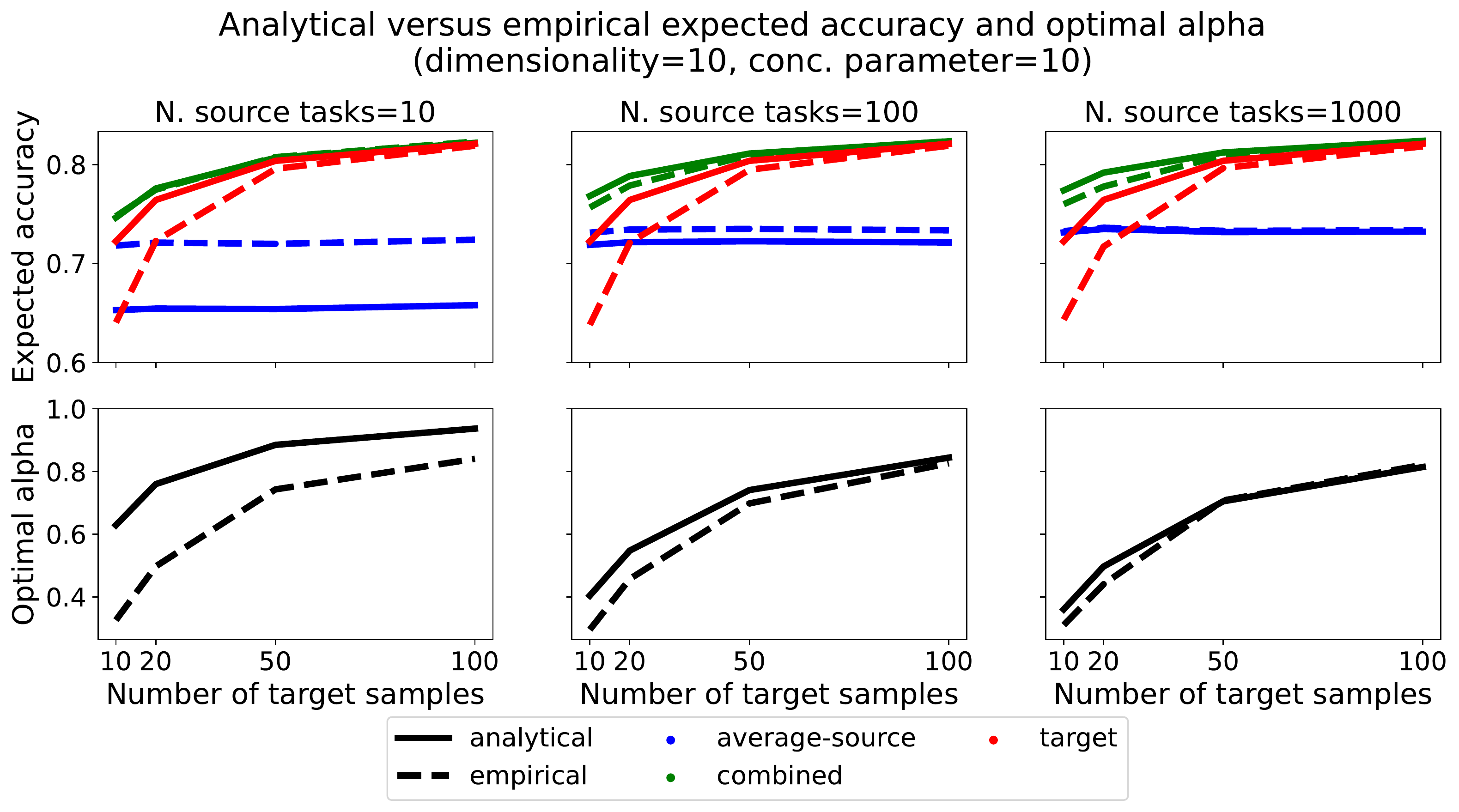}
    \caption{Validating our proposed approximation by comparing the approximated analytical accuracies and empirical accuracies and optimal convex coefficients $\alpha^\ast$ for different amounts of target training data $ n $ and number of source tasks $ J $. 
    }
    \label{fig:analytical-versus-empirical}
\end{figure*}

To validate our approximation we assume that the target class covariance and class 1 mean are known and fix $ d = 10 $, $ \kappa = 10 $. We vary the amount of data available from the target task $ n \in \{10, 20, 50, 100\} $ and the number of source tasks $ J \in \{10, 100, 1000\} $. For each setting we report the average accuracy and average optimal $ \alpha $ from $ 1,000 $ different $ (\nu^{(0)}, \{\omega^{(j)}\}_{j=1}^{J}) $ samples. We report the approximated expected accuracy and optimal $ \alpha$ as calculated using the expression derived in Section \ref{subsec:approximating-optimality}, referred to as the ``analytical" methods in Figure \ref{fig:analytical-versus-empirical}. We also report the accuracy of each classifier on $ 10,000 $ samples from the target task and the corresponding optimal $ \alpha$, referred to as the ``empirical" methods. The empirical accuracies represent the true-but-analytically intractable accuracy. For the analytical combined method we use 100 samples from $ \mathcal{N}\left(\mu_{\omega_{\alpha}}, \Sigma_{\omega_{\alpha}}\right) $ to calculate the risk for each $ \alpha \in \{0, 0.1, 0.2, \hdots, 1.0\} $. 



The gap between the analytical and empirical accuracies associated with the target classifier decreases as the number of samples from the target distribution increases, as seen in each figure in the top row of Figure \ref{fig:analytical-versus-empirical}. This gap in the early part of the regime is caused by the mismatch between asymptotic approximation of the variance associated with the target data. Unsurprisingly, the approximation is better for larger $ n $. Even with the low quality of the approximation for small $ n $, the optimal classifier is able to outperform the target classifier for all $ n $ and the analytical and empirical accuracies are indistinguishable for large $ n $.

Now looking from the left to the right of Figure \ref{fig:analytical-versus-empirical}, we see that the gap between the analytical and empirical risks associated with the average-source and optimal classifiers decreases as we increase the number of source tasks. 
For example, the difference between the empirical and analytical accuracies associated with the average-source task for $ J = 10 $ is quite noticeable whereas the difference for $ J = 1,000 $ is negligible. 
As with the discrepancy for the performance of the target classifier, this is caused by the normal distribution poorly approximating the distribution of the average-source vector for small $ J $.

The validity of our approximation as $ n $ gets large and $ J $ gets large is apparent when evaluating the differences between the optimal convex coefficients (bottom row) -- for $ J = 10 $ the coefficients are separated for the entire regime, for $ J = 100 $ there is meaningful separating for small $ n $ that goes away for larger $ n $, and for $ J = 1,000 $ the separation is negligible nearly immediately. 
While simulation studies designed to evaluate the proposed method in settings with more complicated covariance structures and in the presence of model misspecifications, among others, are required to fully understand the appropriateness of the proposed approximation, we consider the results in Figure\ref{fig:analytical-versus-empirical} as evidence of the appropriateness in the settings studied here. We leave additional simulation studies to future work.

\subsection{The effect of plug-in estimates, concentration, and dimensionality}

\begin{figure*}[h]
    \centering    \includegraphics[width=\linewidth]{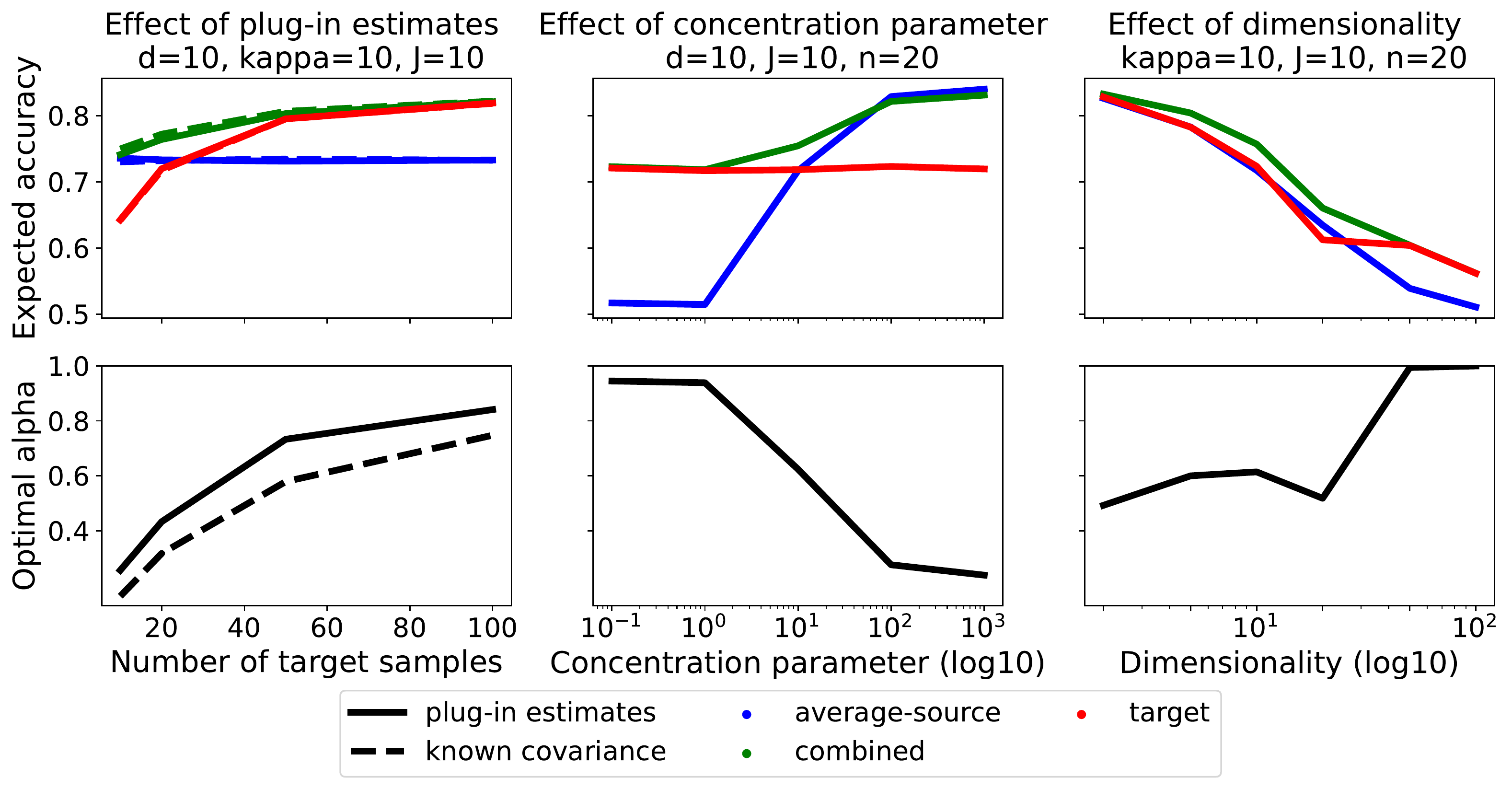}
    \caption{Studying the effect of using plug-in estimates (left) and the effect of varying different generative model parameters (center, right) on the expected accuracy of the average-source, target, and optimal classifiers and on the optimal convex coefficient.}
    \label{fig:effect-of-parameters}
\end{figure*}


Figure \ref{fig:effect-of-parameters} shows the effect of estimating the covariance structure (left column), the effect of different vMF concentration parameters (middle column), and the effect of dimensionality (right column) on the accuracies of the average-source, target, and optimal classifiers and the calculated optimal convex coefficients. 
Unless otherwise stated, we fix the $ d=10 $, $ J=100 $, $ \kappa = 10 $ and $ n=20 $. The classifiers are evaluated using 10,000 samples from the target distribution.

The left column of Figure \ref{fig:effect-of-parameters} illustrates the effect of estimating the target task's class conditional covariance structure $ \Sigma^{(0)} $ and class 1 conditional mean $ \nu^{(0)} $ and using these estimates as plug-ins for their population values when approximating the risk described in Eq. \ref{eq:risk-integral}. 
In particular, we compare the expected accuracy and optimal coefficient when using the plug-in estimates (solid lines) $ \hat{\Sigma}^{(0)} $ and $ \hat{\nu}^{(0)} $ to using the population covariance $ \Sigma^{(0)} $ and $ {\nu}^{(0)} $ (dashed lines).
We note that the difference between the performance of the optimal classifiers in the two paradigms is smaller than the difference between the performance of optimal classifier and the target classifier for small $ n $. 
This behavior is expected, as the optimal classifier has access to more information through the average-source projection vector. 
Finally, we note that the difference between the two optimal coefficients is smaller for the poles of the regime and larger in the middle. We think that this is due to higher entropy states between wanting to use the ``high bias, low variance" average-source classifier and the ``low bias, high variance" target classifier. 

For both the middle and right columns of Figure \ref{fig:effect-of-parameters} we study only the plug-in classifiers.
The middle column of Figure \ref{fig:effect-of-parameters} investigates the effect of the vMF concentration parameter $ \kappa $. 
Recall that as  $\kappa $ gets larger the expected cosine distance between samples from the vMF distribution gets smaller. This means that the expected cosine distance between the average-source projection vector and the true-but-unknown target projection gets smaller. 
Indeed, through the expected accuracies of the average-source and target classifiers we see that the average-source classifier dominates the target classifier in the latter part of the studied regime due to the average-source vector providing good bias. 
Notably, the combined classifier is always as effective and sometimes better than the target classifier but is slightly less effective than the average-source classifier when $ \kappa $ is large. This, again, is due to the appropriateness of modeling the average-source vector as Gaussian.
The optimal convex coefficient is close to 1 when the vMF distribution is close to the uniform distribution on the unit sphere ($ \kappa $ small) and closer to 0 when vMF distribution is closer to a point mass ($\kappa $ large).

The right column of Figure \ref{fig:effect-of-parameters} shows the effect of the dimensionality of the classification problem on the expected accuracies and optimal coefficient. The top figure demonstrates that the optimal classifier is always as good as and sometimes better than both the average-source and target classifiers, with the margin being small when the dimensionality is both small and large. The reason the margin between the accuracies starts small, gets larger, and then becomes small again is likely due to the interplay between the estimation error associated with covariance structure and the relative concentration of the source vectors. We do not investigate this complicated interplay further. The optimal coefficient gets progressively larger as the dimensionality increases with the exception of a dip at $ d=20 $. We think this dip is due to a regime change in the interplay mentioned previously.


\section{Applications to physiological prediction problems}

We next study the proposed class of classifiers in the context of three physiological prediction problems: EEG-based cognitive load classification, EEG-based stress classification, and ECG-based social stress classification. 
Each of these problems has large distributional variability across persons, devices, sessions, and tasks.
Moreover, labeled data in these tasks is expensive -- non-overlapping feature vectors can require up to 45 seconds of recording to obtain. That is, large improvements in classification metrics near the \textit{beginning} of the in-task data regime is important in mitigating the amount of time required for a Human-Computer interface to produce relevant predictions and is thus necessary for making these types of devices usable. 

The dataset related to EEG-based cognitive load classification task is proprietary. We include the results because there is a (relatively) large number of participants with multiple sessions per participant and the cognitive load task is a representative high-level cognitive state classification problem.
Both the EEG-based\citep{zyma2019electroencephalograms} and ECG-based stress \citep{schmidt2018introducing} classification are publicly available. Given the complicated nature of physiological prediction problems, previous works that use these datasets typically choose an arbitrary amount of training data for each session, train a model, and report classification metrics related to a held out test set (e.g., \citep{electronics10091079} (EEG) and \citep{indikawati2020stress} (ECG)) or held out participants (e.g., \citep{9726225} (EEG) and \citep{10.1145/3397482.3450732, gil2022human} (ECG)). 
Our focus, while similar, is fundamentally different: we are interested in classification metrics as a function of amount of training data seen.

In each setting we have access to a small amount of data from a target study participant and the projection vectors from other participants. The data for each subject is processed such that the assumptions of Eq. \eqref{eq:model-assumptions} are matched as closely as possible. For example, we use the available training data from the target participant to force the class conditional means to be on the unit sphere and for their midpoint to cross through the origin. Further, we normalize the learned projection vectors so that the assumption that the vectors come from a von-Mises Fisher distribution is sensible. 

The descriptions of the cognitive load and stress datasets are altered versions of the descriptions found in Chen et al. \citep{10.3389/fnhum.2022.930291}. 
Unless otherwise stated, the balanced accuracy and the convex coefficient corresponding to each method are calculated using 100 different train-test splits for each participant. 
Conditioned on the class-type, the windowed data used for training are consecutive windows.
A grid search in $ \{0, 0.1, 0.2, \hdots, 1.0\} $ was used when calculating convex coefficients.

\subsection{Cognitive load (EEG)}
\label{subsec:matb}

\begin{figure*}[h]
    \captionsetup[subfigure]{justification=centering}
    \begin{subfigure}{\textwidth}
    \centering
    \includegraphics[width=\linewidth]{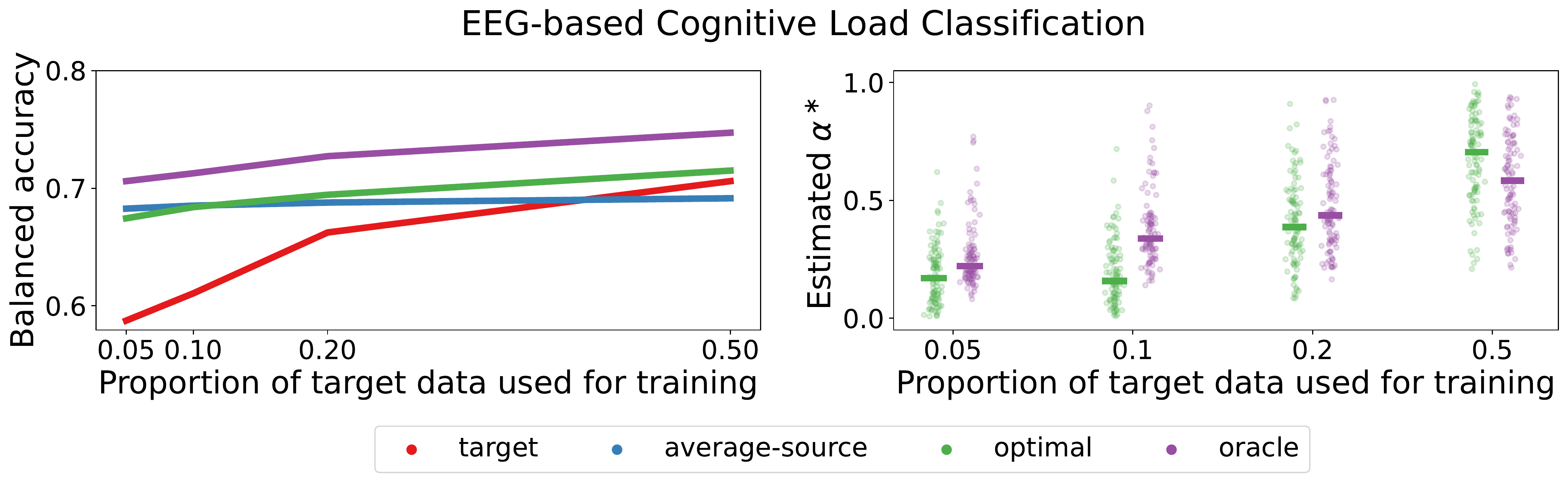}
    \end{subfigure} \\
    \begin{subfigure}{\textwidth}
          \centering
          \includegraphics[width=\linewidth]{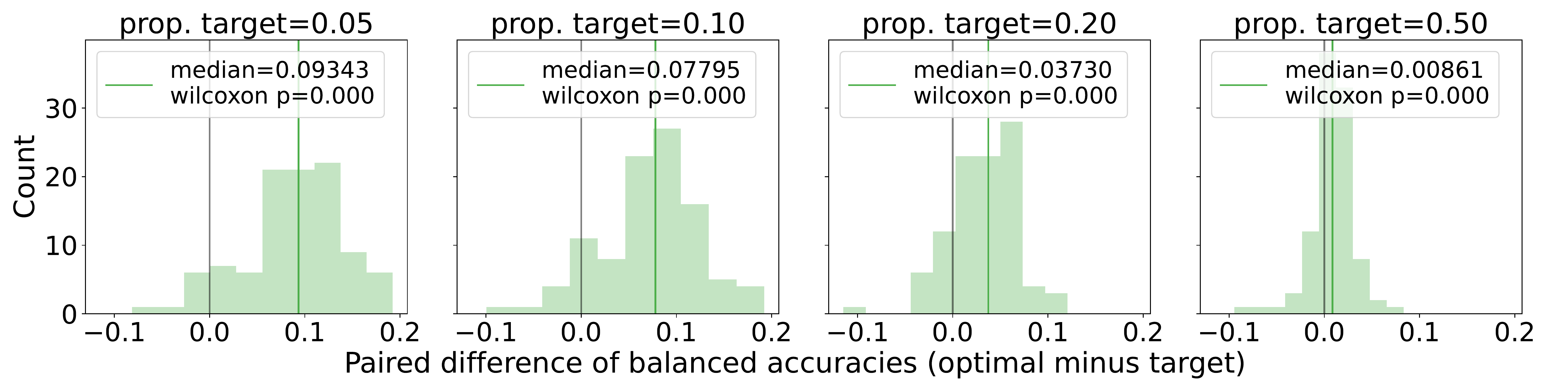}  
          \label{subfig:matb-histograms}
    \end{subfigure}
    \caption{Balanced accuracy and relevant convex coefficients (top) and relative performance of the optimal and target classifiers (bottom) for the MATB-II cognitive load classification task.
    }
\label{fig:cognitive-load}
\end{figure*}

The first dataset we consider was collected under NASA's Multi-Attribute Task Battery II (MATB-II) protocol. MATB-II is used to understand a pilot's ability to perform under various cognitive load requirements \citep{santiago2011multi} by attempting to induce four different levels of cognitive load -- no (passive), low, medium, and high ---that are a function of how many tasks the participant must actively tend to.

The data includes 50 healthy subjects with normal or corrected-to-normal vision. There were 29 female and 21 male participants and each participant was between the ages of 18 and 39 (mean 25.9, std 5.4 years). Each participant was familiarized with MATB-II and then participated in two sessions containing three segments. The three segments were further divided into blocks with the four different levels of cognitive requirements. The sessions lasted around 50 minutes and were separated by a 10 minute break. We focus our analysis on a per-subject basis, meaning there will be two sessions per subject for a total of 100 different sessions.

The EEG data was recorded using a 24-channel Smarting MOBI device and was processed using high pass (0.5 Hz) and low pass (30 Hz) filters and segmented in ten second, non-overlapping windows. Once the EEG-data was windowed we calculated the mass in the frequency domain for the theta (4-8 Hz), alpha (8-12 Hz), and lower beta (12-20 Hz) bands. We then normalized the mass of each band on a per channel basis. In our analysis we consider only the frontal channels \{Fp1, Fp2, F3 F4, F7, F8, Fz, aFz\}. Our choice in channels and bands is an attempt to mitigate the number of features while maintaining the presence of known cognitive load indicators \citep{article}. The results reported in Figure \ref{fig:cognitive-load} are for this $ (3\times 8) = 24 $-dimensional two class problem \{no \& low cognitive load, medium \& high cognitive load\}.

For a fixed session we randomly sample a continuous proportion of the participant's windowed data $ p \in \{0.05, 0.1, 0.2, 0.5\} $ and also have access to the projection vectors corresponding to all sessions except for the target participant's other session (i.e., we have 100 - 1 - 1 = 98 source projection vectors). As mentioned above, we use the training data to learn a translation and scaling to best match the model assumptions of Section \ref{sec:generative-model}. 

The top left figure of Figure \ref{fig:cognitive-load} shows the mean balanced accuracy on the non-sampled windows of four different classifiers: the average-source classifier, the target classifier, the optimal classifier, and the oracle classifier. The average-source, target, and optimal classifiers are as described in Section \ref{subsec:simulations}. The oracle classifier is the convex combination of the average-source and target projection vectors that performs the best on the held out test set. The median balanced accuracy of each classifier is the median (across sessions) calculated from the mean balanced accuracy of 100 different train-test samplings for each session.

The relative behaviors of the average-source, target and optimal classifiers in this experiment are similar to what we observe when varying the amount of target data in the simulations for large $ \kappa $ -- the average-source classifier outperforms the target classifier in small data regimes, the target classifier outperforms the average-source classifier in large data regimes, and the optimal classifier is able to outperform or match the performance of both classifiers throughout the regime. Indeed, in this experiment the empirical value of $ \kappa $ when estimating the projection vectors using all of each session's data is approximately 17.2.

The top right figure of Figure \ref{fig:cognitive-load} shows scatter plots of the convex coefficients for the optimal and oracle methods. Each dot represents the average of 100 coefficients for a particular session for a given proportion of training data from the target task (i.e., one dot per session). The median coefficient is represented by a short line segment. The median coefficient for both the oracle and the optimal classifiers get closer to 1 as more target data is available. This behavior is intuitive, as we'd expect the optimal algorithm to favor the in-distribution data when the estimated variance of the target classifier is ``small".

The bottom row of Figure \ref{fig:cognitive-load} is the set of histograms of the difference between the optimal classifier's balanced accuracy and the target classifier's balanced accuracy where each count represents a single session. These histograms give us a better sense of the relative performance of the two classifiers -- a distribution centered around 0 would mean that we have no reason to prefer the optimal classifier over the target classifier and where a distribution shifted to the right of 0 would mean that we would prefer the optimal classifier to the target classifier.

For $ p=0.05 $ the optimal classifier outperforms the target classifier for 92 of the 100 sessions with differences as large as 19.2\% and a median absolute accuracy improvement of about 9.3\%. 
The story is similarly dramatic for $ p = 0.10 $ with the optimal classifier outperforming the target classifier for 92 of the 100 sessions, a maximum difference of about 19.2\%, and a median difference of 7.8\%. 
For $ p = 0.2 $ the distribution of the differences is still shifted to the right of $ 0 $ with a non-trivial median absolute improvement of about 3.7\%, a maximum improvement of 12\%, and an improvement for 81 of the sessions.. 
For $ p = 0.5 $ the optimal classifier outperforms the target classifier for 76 of the 100 sessions, though the distribution is only slightly shifted to the right of $ 0 $. 
The p-values, up to 3 decimal places, from the one-sided Wilcoxon's rank-sum test for the hypothesis that the distribution of the paired differences is symmetric and centered around 0 are less than $ 0.001 $ for each proportion of available target data that we considered.

\subsection{Stress from mental math (EEG)}
\label{subsec:mental-math}

\begin{figure*}[h]
    \captionsetup[subfigure]{justification=centering}
    \begin{subfigure}{\textwidth}
    \centering
    \includegraphics[width=\linewidth]{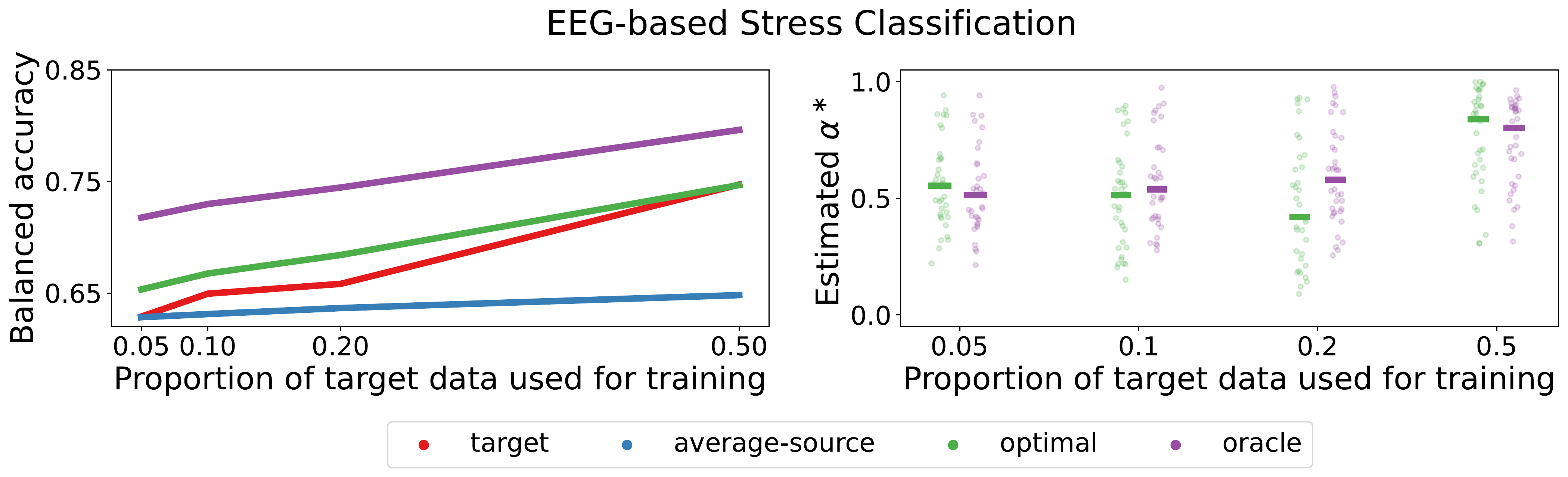}
    \end{subfigure} \\
    \begin{subfigure}{\textwidth}
          \centering
          \includegraphics[width=\linewidth]{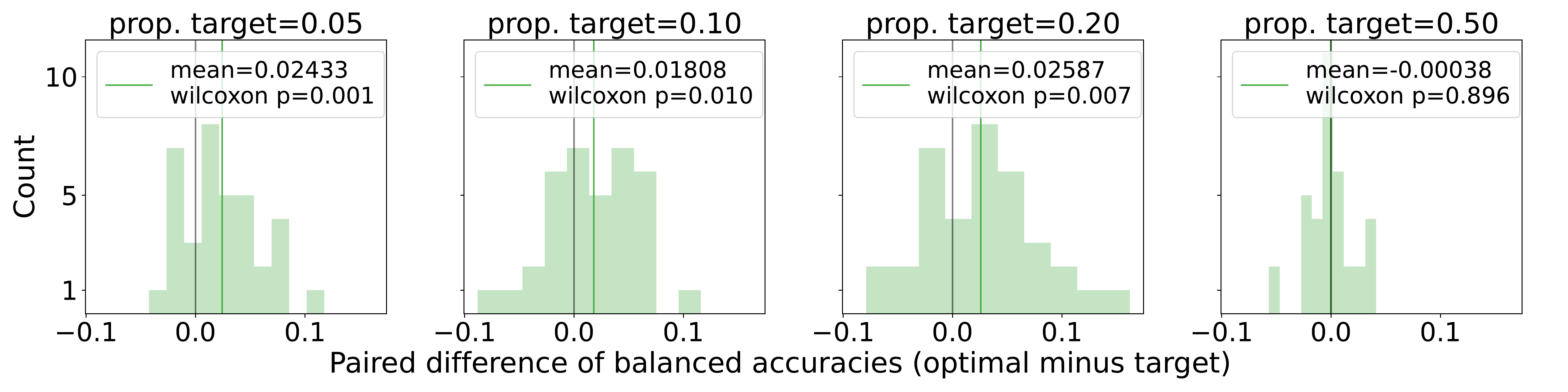}  
          \label{subfig:mm-histograms}
    \end{subfigure}
    \caption{Balanced accuracy and relevant convex coefficients (top) and relative performance of the optimal and target classifiers on a per-participant basis (bottom) for the Mental Math EEG-based stress classification task.
    }
\label{fig:stress}
\end{figure*}

In the next study we consider there are two recordings for each session -- one corresponding to a resting state and one corresponding to a stressed state. 
For the resting state, participants counted mentally (i.e., without speaking or moving their fingers) with their eyes closed for three minutes. 
For the stressful state, participants were given a four digit number (e.g., 1253) and a two digit number (e.g., 43) and asked to recursively subtract the two digit number from the four digit number for 4 minutes. 
This type of mental arithmetic is known to induce stress \citep{noto2005relationship}.

There were initially 66 participants (47 women and 19 men) of matched age in the study. 30 of the participants were excluded from the released data due to poor EEG quality.
Thus we consider the provided set of 36 participants first analyzed by the study's authors \citep{zyma2019electroencephalograms}. 
The released EEG data was preprocessed via a high-pass filter and a power line notch filter (50 Hz). Artifacts such as eye movements and muscle tension were removed via ICA. 
We windowed the data into two and a half second chunks with no overlap and consider the two-class classification task \{stressed, not stressed\} with access only to the channels along the centerline \{Fz, Cz, Pz\} and the theta, alpha and lower beta bands described above. 
The results of this experiment are displayed in Figure \ref{fig:stress} and are structured in the same way as the cognitive load results. 

For this study we see relative parity between the target and average-source classifiers when $ p = 0.05 $. In this case, the optimal classifier is able to leverage the discriminative information in both sets of information and improve the balanced accuracy. This win is maintained until the target classifier performance matches the optimal classifier performance for $ p =0.5 $. The poor performance of the average-source classifier is likely due to the empirical value for $ \kappa $ being less than 3. 

Interestingly, we do not see as clear of a trend for the median convex coefficients in the top right figure. They are relatively stagnant between $ p=0.05, 0.1 $ and $ 0.2 $ before jumping considerably closer to $ 1 $ for $ p=0.5 $.

When comparing the optimal classifier to the target classifier on a per-participant basis directly (bottom row) it is clear that the optimal classifier is favorable: for $ p = 0.05, 0.10 $ and $ p=0.2 $ the optimal classifier outperforms the target classifier for 25, 24, and 24 of the 36 participants, respectively, and the median absolute difference of these wins is in the 1.8\% - 2.6\% range for all three settings with maximum improvements of 19.2 for $ p=0.05, $  19.2 for $ p=0.1, $ and 12.1 for $ p=0.2$. As with the cognitive load task, this narrative shifts for $ p=0.5 $ as the distribution of the differences is approximately centered around 0. The p-values from the one-sided rank-sum test reflect these observations: 0.001, 0.01, 0.007, and 0.896 for $ p =0.05, 0.1, 0.2$, and $ 0.5 $, respectively.

\subsection{Stress in social settings (ECG)}
\label{subsec:wesad}

\begin{figure*}[h]
    \captionsetup[subfigure]{justification=centering}
    \begin{subfigure}{\textwidth}
        \centering
        \includegraphics[width=\linewidth]{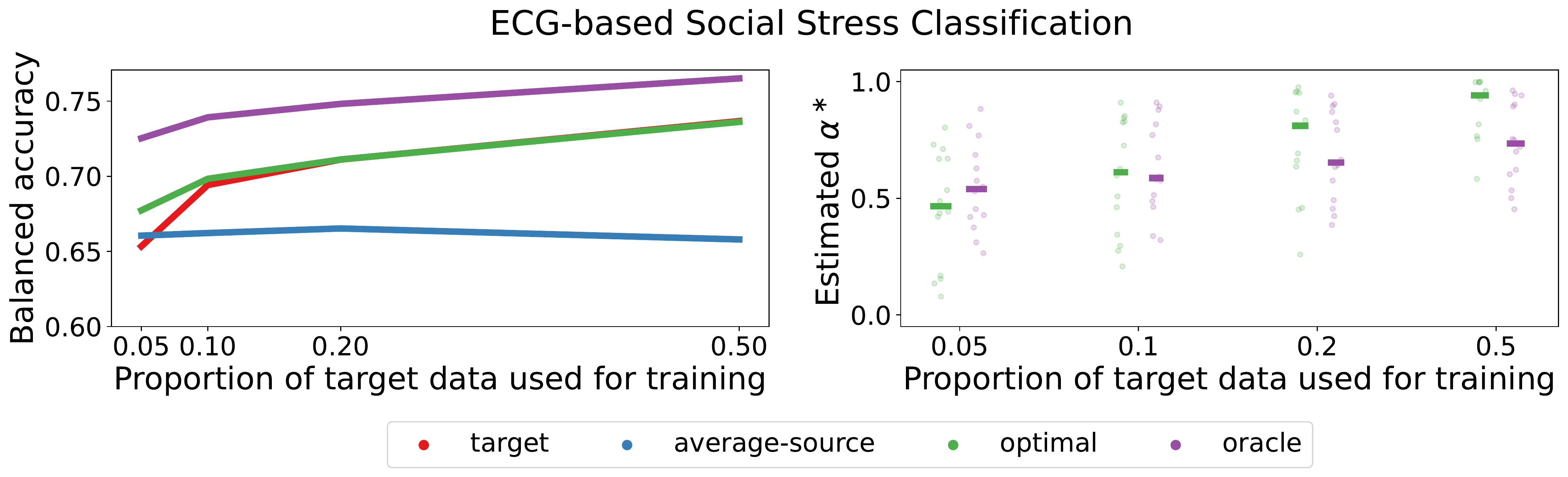}
    \end{subfigure} \\
    \begin{subfigure}{\textwidth}
        \centering
        \includegraphics[width=\linewidth]{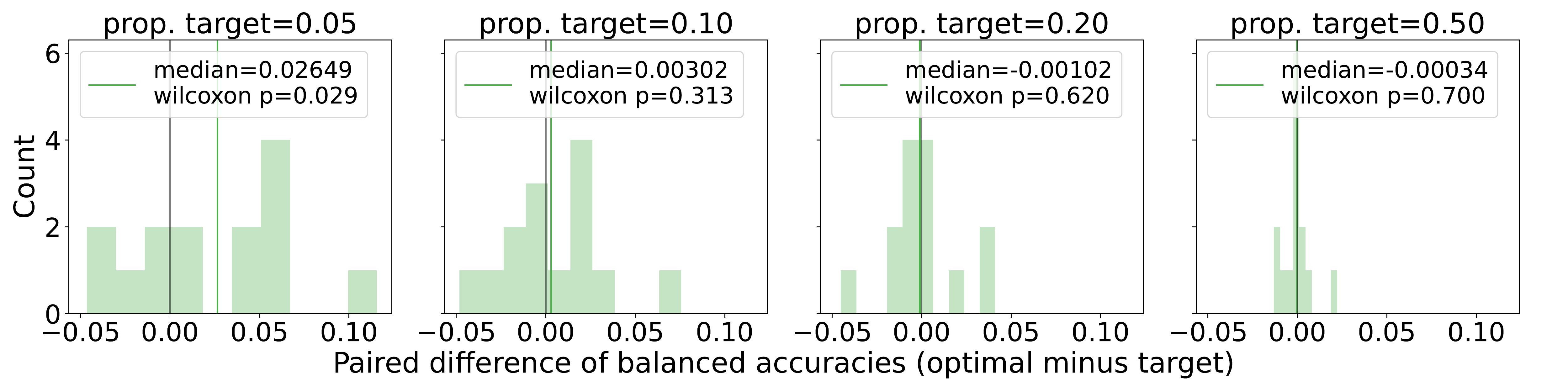}  
        \label{subfig:ecg-histograms}
    \end{subfigure}
    \caption{Balanced accuracy and relevant convex coefficients (top) and relative performance of the optimal and target classifiers on a per-participant basis (bottom) for the Social Stress, ECG-based  classification task.
    }
\label{fig:ecg}
\end{figure*}

The last dataset we consider is the WEarable Stress and Affect Detection (WESAD) dataset \citep{schmidt2018introducing}. For WESAD, the researchers collected multi-modal data while participants underwent a neutral baseline condition, an amusement condition and a stress condition. The participants meditated between conditions. For our purposes, we will only consider the baseline condition where participants passively read a neutral magazine for approximately 20 minutes and the stress condition where participants went through a combination of the Trier Social Stress Test and a mental arithmetic task for a total of 10 minutes. 

For our analysis, we consider 14 of the 15 participants and only work with their corresponding ECG data recorded at 700 Hz. Before featurizing the data, we first downsampled to 100 Hz and split the time series into 15 second, non-overlapping windows. We used Hamilton's peak detection algorithm \citep{4122227} to find the time between heartbeats for a given window. We then calculated the proportion of intervals larger than 20 milliseconds, the normalized standard deviation of the interval length, and the ratio of the high (between 15 and 40 hz) and low (between 4 and 15 hz) frequencies of the interval waveform after applying a Lomb-Scargle correction for waves with uneven sampling. These three features are known to have discriminative power in the context of stress prediction \citep{hrv-review}, though typically for larger time windows.

We report the same metrics for this dataset in Figure \ref{fig:ecg} as we do for the two EEG studies above: the mean balanced accuracies are given in the top left figure, the convex coefficients for the optimal and oracle classifiers are given in the top right and the paired difference histograms between the optimal classifier's balanced accuracy and the target classifier's balanced accuracy are given in the bottom row.

The relative behaviors of the classifiers in this study is similar to the behaviors in EEG-based stress study above. The optimal classifier is able to outperform the other two classifiers for $ p = 0.05 $ and is matched by the target classifier for the rest of the regime. The average-source classifier is never preferred and the empirical value of $ \kappa $ is approximately 1.5. The distributions of the optimal coefficients get closer to 1 as $ p $ increases but are considerably higher compared to the MATB study for each value of $ p $ -- likely due to the large difference between the empirical values of $ \kappa $ across the two problems.

Lastly, the paired difference histograms for $ p = 0.05 $ favors the optimal classifier. The histograms for $ p=0.1, 0.2, $ and $ 0.5 $ are inconclusive. The p-values for Wilcoxon's rank-sum test are $ 0.029, 0.313, 0.620 $ and $ 0.700 $ for $ p = 0.05, 0.1, 0.2 $ and $ 0.5 $, respectively. 

\subsection{Visualizing the projection vectors}
\label{subsec:visualizations}

\begin{figure*}[h]
    \captionsetup[subfigure]{justification=centering}
    \begin{subfigure}{0.45\textwidth}
    \centering
    \includegraphics[width=\linewidth]{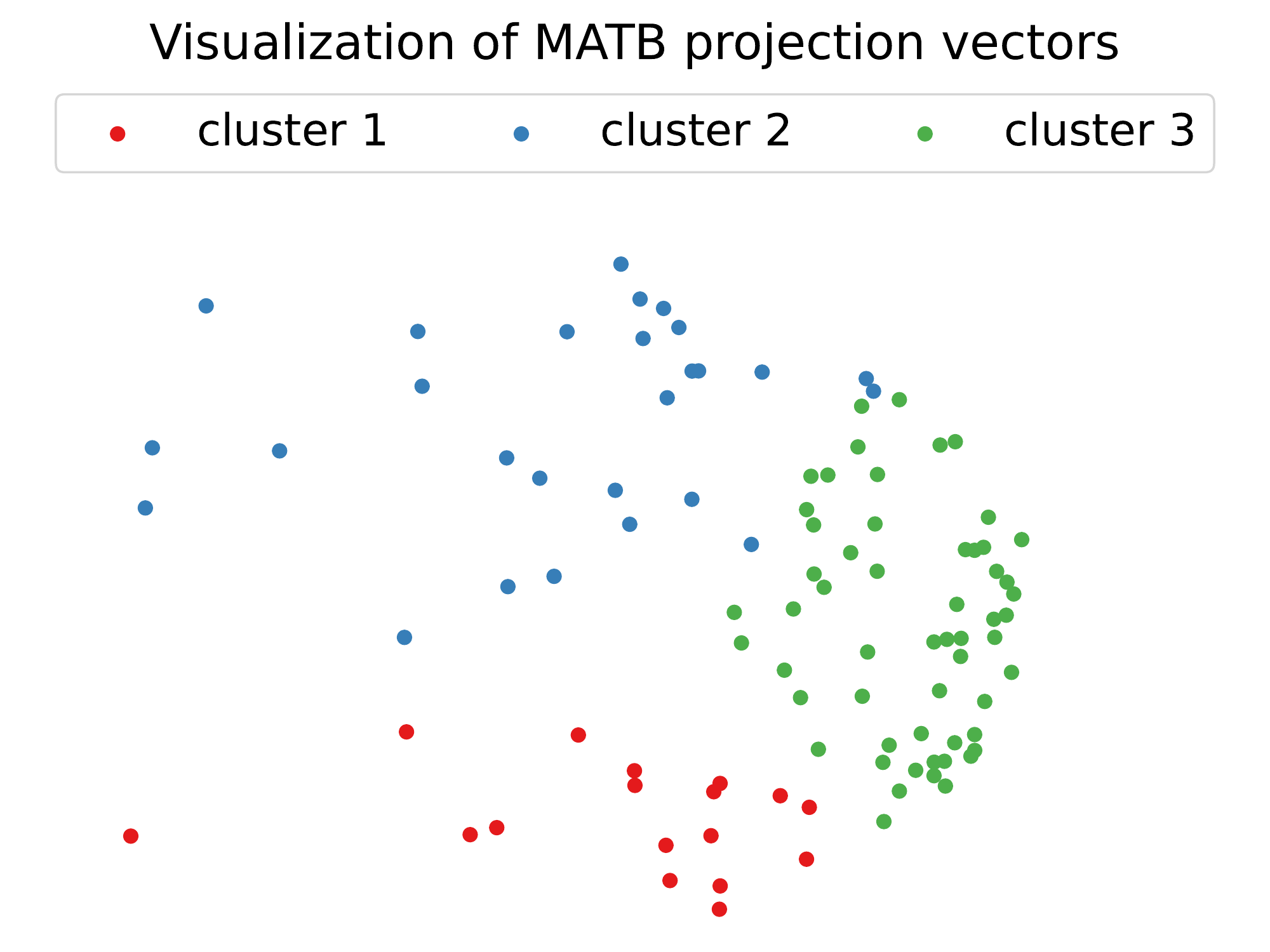}
    \end{subfigure}
    \begin{subfigure}{0.45\textwidth}
          \centering
          \includegraphics[width=\linewidth]{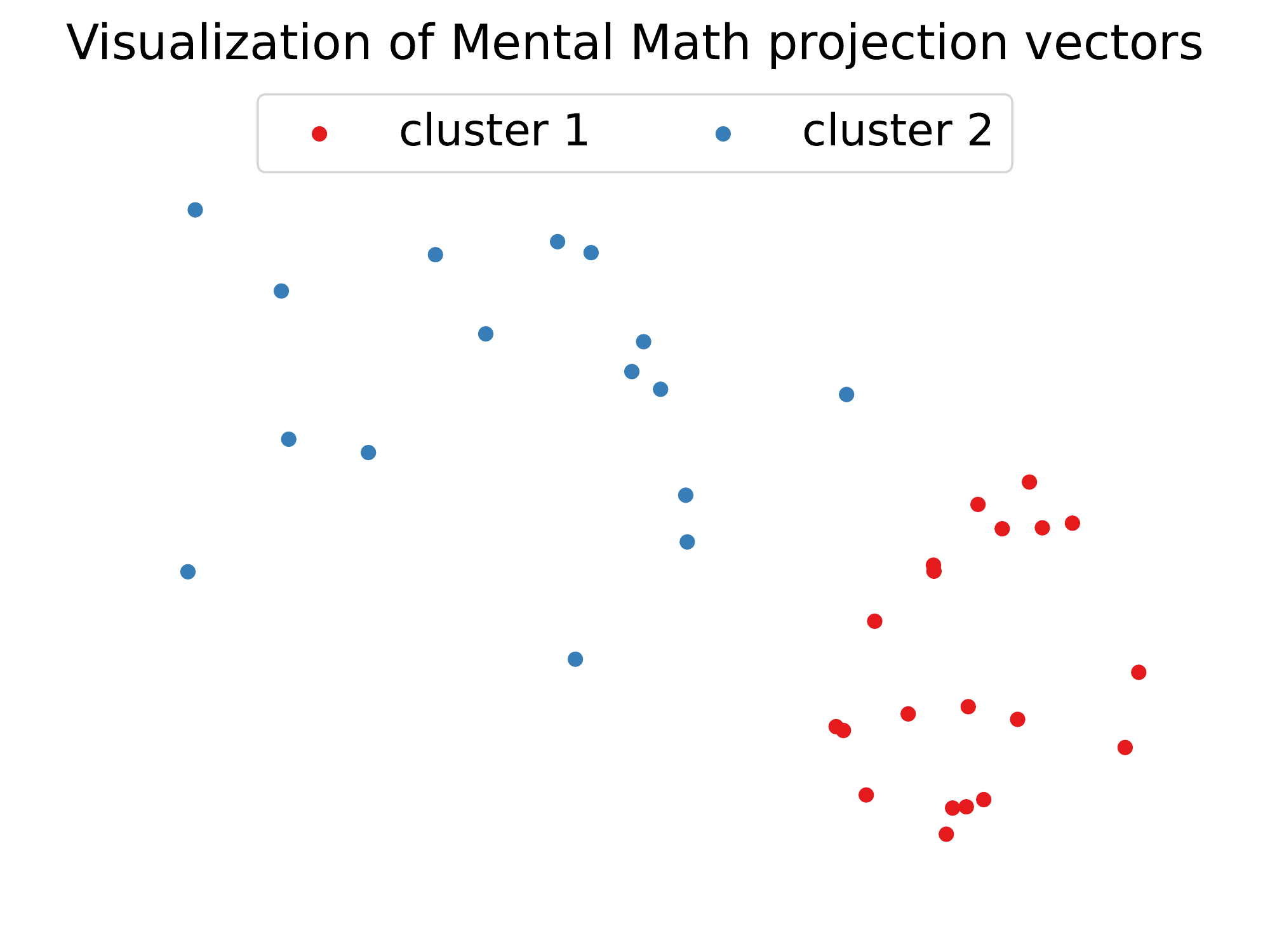}  
    \end{subfigure}
    \\
    
    \begin{subfigure}{\textwidth}  
    \centering
    \includegraphics[width=0.45\linewidth]{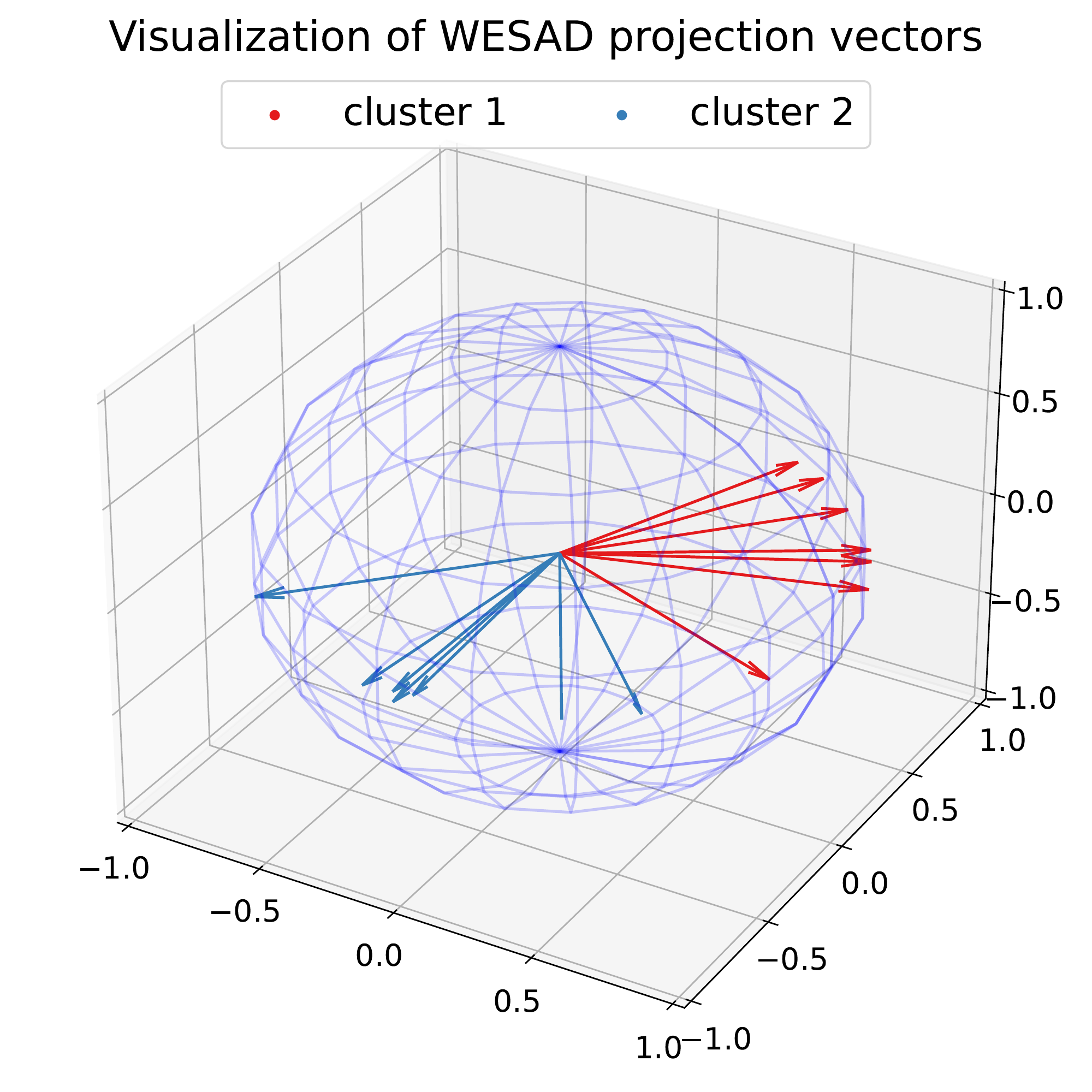}
    \end{subfigure}    
    \caption{Visualizations of the projection vectors for each of the three datasets under study where each dot or arrow corresponds to a session. 
    The projection vectors were estimated using the entire data from each session.
    For the MATB (top left) and Mental Math (top right) visualizations we show the first two principal components scaled by their corresponding eigenvalues of the $ J \times J $ cosine similarity matrix. 
    The WESAD visualization (bottom) shows the three dimensional projection vectors.
    Colors denote the component of a Gaussian mixture model fitted to the projection vectors. 
    }
\label{fig:visualization}
\end{figure*}

The classification results above provide evidence that our proposed approximation to the optimal combination of the average-source and target projection vectors is useful from the perspective of improving the balanced accuracy. There is, however, a consistent gap that remains between the performance of the optimal classifier and the performance of the oracle classifier. To begin to diagnose potential issues with our model, we visualize the projection vectors from each of the tasks. 

The three subfigures of Figure \ref{fig:visualization} show representations of the projection vectors for each task. The dots in the top row correspond to projection vectors from sessions from the MATB dataset (left) and the Mental Math dataset (right). The arrows with endpoints on the sphere in the bottom row correspond to projection vectors from sessions from WESAD. For these visualizations the entire dataset was used to estimate the projection vectors. The two-dimensional representations for MATB and Mental Math are first two components of the spectral embedding \citep{ase} of the affinity matrix $ A $ with entries $ a_{ij} = (\omega^{(i)\top}\omega^{(j)} + 1) / 2 $ and $ a_{ii} = 0 $. The projection vectors for the WESAD task are three dimensional and are thus amenable to visualization. 

For each task we clustered the representations of the projection vectors using a Gaussian mixture model where the number of components was automatically selected via minimization of the Bayesian Information Criterion (BIC). The colors of the dots and arrows reflect this cluster membership.
The BIC objective function prefers a model with at least two components to a model with a single component for all of the classification problems -- meaning that modeling the distribution of the source vectors as a unimodal von-Mises Fisher distribution is likely wrong and that a multi-modal von-Mises Fisher distribution may be more appropriate. We do not pursue this idea further but do think that it is could be a fruitful future research direction if trying to mitigate the gap between the performances of the optimal and oracle classifiers.

\subsection{The effect of the number of samples used to calculate $\alpha^{*}$}
\begin{figure}
    \centering
    \includegraphics[width=0.6\linewidth]{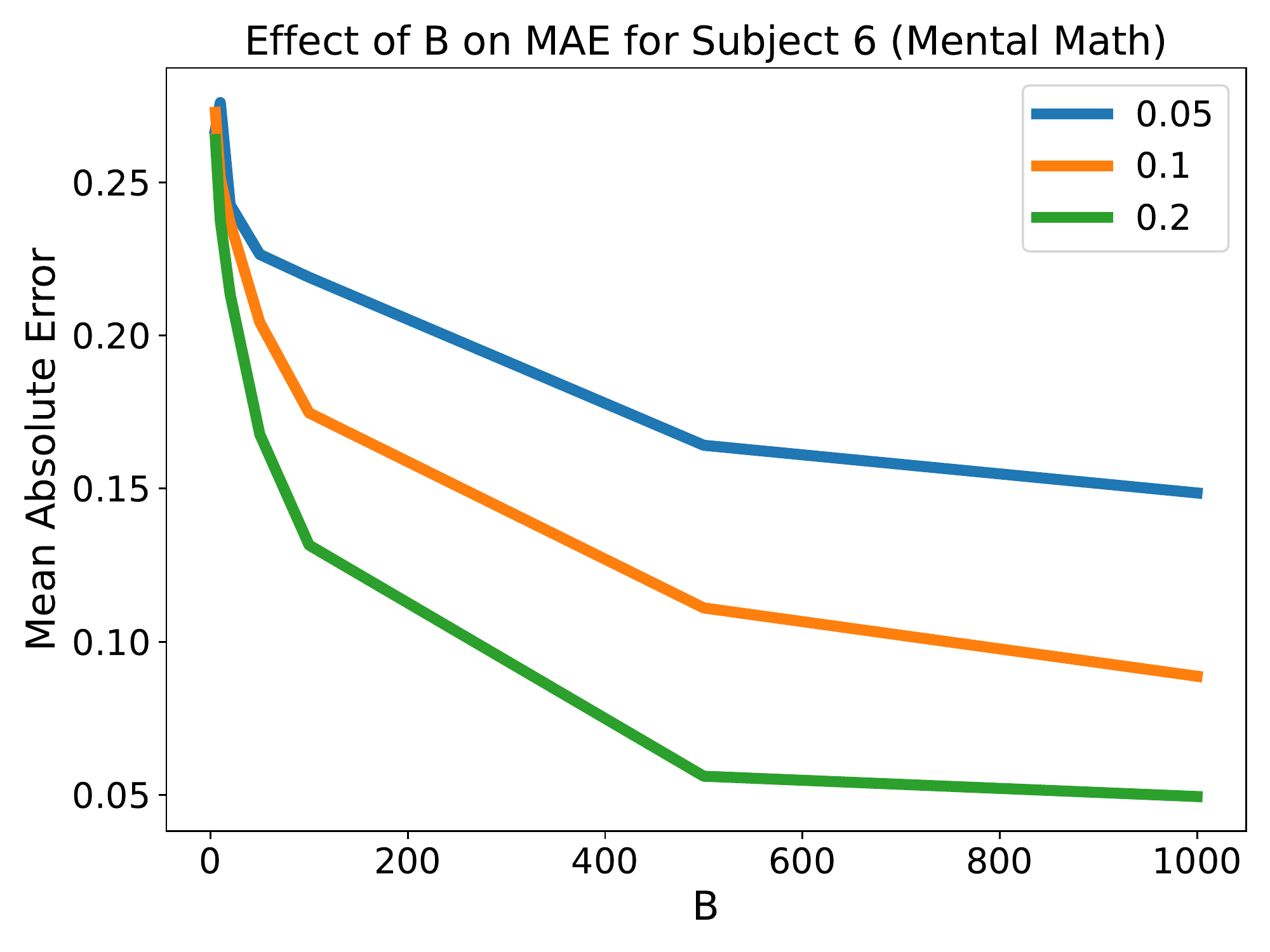}
    \caption{The effect of the number samples sampled from the distribution of $ \omega_{\alpha} $ on the absolute error between the optimal $ \alpha $ calculated using $ B $ samples and the optimal $ \alpha $ calculated using $ B^{*}=10,000 $ samples for subject 6 of the Mental Math dataset. }
    \label{fig:effect-of-B}
\end{figure}

In the simulation experiments described in Section \ref{subsec:simulations} and the applications to different physiological prediction problems in Sections \ref{subsec:matb}, \ref{subsec:mental-math}, and \ref{subsec:wesad} we used 100 samples from the distribution of $ \omega_{\alpha} $ to estimate the risk for a given $ \alpha $. There is no way to know  \textit{a priori} how many samples is sufficient for estimating the optimal coefficient. We can, however, study how different amounts of samples effect the absolute error of the optimal coefficient compared to an coefficient calculated using an unrealistic amount of samples. For this analysis we focus on a single session from the Mental Math dataset described in \ref{subsec:mental-math}. The dataset choice was a bit arbitrary. The session was chosen because it is the session where the optimal classifier performs closest to the median balanced accuracy for $ p = 0.1 $.

Figure \ref{fig:effect-of-B} shows the effect of $ B $, the number of samples from the distribution of $ \omega_{\alpha} $ used to calculate the risk for a given $ \alpha $, on the mean absolute error when compared to a convex coefficient calculated using $B^{*} = 10,000 $ samples. The mean absolute errors shown are calculated for $ p \in \{0.05, 0.1, 0.2\} $ by first sampling a proportion of data $ p $ from the target task, training the target classifier using the sampled data, and then estimating the optimal coefficient using $ B^{*} = 10,000 $ samples from the distribution of $ \omega_{\alpha} $. We then compare this optimal coefficient to coefficient found using $ B \in \{5, 10, 20, 50, 100, 200, 500, 1000\} $ samples from the distribution of $ \omega_{\alpha} $ 30 different times, calculate the absolute difference, and record the mean. The lines shown in Figure \ref{fig:effect-of-B} is the average of 100 different training sets. In this experiment the coefficients $ \alpha \in \{0, 0.1, \hdots, 1.0\} $ were evaluated. 

There are a few things of note. First, when there is more target data available the fewer samples from the distribution of $ \omega_{\alpha} $ are needed to obtain a specific value of mean absolute error. Second, the mean absolute error curves appear to be a negative exponential function of $ B $ and, for this subject, it seems that the benefit of more samples decays quite quickly after $ B = 500 $. Lastly, though the closer the convex coefficients are to the coefficient calculated using $ B^{*} $ samples the more closely the classifier will perform to the analytically derived optimal classifier, the gap between the performance of the oracle classifier and the optimal classifier in the real data sections above indicates that there may be some benefit from a non-zero mean absolute error.

\subsection{Computational complexity}
Assuming that we have access to the source projection vectors $ \omega^{(1)}, \hdots, \omega^{(J)} $ and target data $ \{X_{i}^{(0)}, Y_{i}^{(0)}\}_{i=1}^{n} $ and letting $ B $ be the number of bootstrap samples and $ h $ be the number of evaluated classifiers in $ \mc{H} $, the computational complexities for obtaining the projection vectors associated with the three algorithms studied above are as follows: the average-source classifier is $ O(J\cdot d) $; the target classifier is $ O(n\cdot d \cdot \text{min}(n, d) + \text{min}(n, d)^{3}) $; and the approximately optimal classifier is $ O(J\cdot d + n\cdot d \cdot \text{min}(n, d) + \text{min}(n, d)^{3} + B^{2}\cdot h) $. That is, using the approximately optimal classifier incurs an additional computational cost that is quadratic in $ B $ and linear in $ h $ when assuming that sampling from a multivariate Gaussian and evaluating the error for each random sample are both $ O(1) $.

\subsection{Privacy considerations}
\label{subsec:privacy}

As presented in Algorithm \ref{alg:optimal}, the process for calculating the optimal convex coefficient $ \alpha^{*} $ requires access to the normalized source projection vectors $ \{\omega^{(j)}\}_{j=1}^{J} $. This requirement can be prohibitive in applications where the data (or derivatives thereof) from a source task are required to stay local to a single device or are otherwise unable to be shared. For example, it is common for researchers to collect data in a lab setting, deploy a similar data collection protocol in a more realistic setting, and to use the in-lab data as a source task and the real-world data as a target task. Depending on the privacy agreements between the researchers and the subjects, it may be impossible to use the source data directly.

The requirements for Algorithm \ref{alg:optimal} can be changed to address these privacy concerns by calculating the average source vector $ \hat{\mu} $ and its corresponding standard error $ \Psi $ in the lab setting and only sharing these two parameters. Indeed, given $ \hat{\mu} $ and $ \Psi $ the algorithm is independent of the normalized source vectors and can be the only thing stored and shared with devices and systems collecting data from the target task.

\section{Discussion}
The approximation to the optimal convex combination of the target and average-source projection vector proposed in Section \ref{sec:generative-model} is effective in improving the classification performance in simulation and, more importantly, across different physiological prediction settings. The improvement is both operationally significant and statistically significant in settings where very little training data from the target distribution is available. In most Human-Computer Interface (HCI) systems an improvement in this part of the regime is the most critical as manufacturers want to mitigate the amount of configuration time (i.e., the time spent collecting labeled data) the users endure and, more generally, make the systems easier to use. We think that our proposed method, along with its privacy-preserving properties inherent to parameter estimation, is helpful towards that goal.

With that said, there are limitations in our work. For example, the derivation of the optimal convex coefficient and, subsequently, our proposed approximation is only valid for the two class problem. We do not think that an extension to the multi-class problem is trivial, though treating a multi-class problem as multiple instances of the two class problem is a potential way forward \citep{tibshirani, li2006using}. 

Similarly, our choice to use a single coefficient on the average-source projection vector, as opposed to one coefficient per source task, may be limiting in situations where the source vectors are not well concentrated. In the WESAD analysis where $ \kappa \approx 1.5 $, for example, it may be possible to maintain an advantage over the target classifier for a larger section of the regime with a more flexible class of hypotheses. The flexibility, however, comes at the cost of privacy and computational resources. A potential middle ground between maximal flexibility and the combination of privacy preservation and computational costs modeling the distribution of the source projection vectors as a multi-modal vMF where the algorithm would only need access to the mean direction vector and standard errors associated with each constituent distribution. The visualizations in Section \ref{subsec:visualizations} provide evidence that this model may be more appropriate than the one studied here.

\bibliographystyle{iclr2021/iclr2021_conference}
\bibliography{biblio}

\appendix
\appendix

\section{Derivation of the analytical expression for classification error w.r.t target distribution}
\label{sec:app-derivation}

Suppose the target distribution is given by $\mc{P} = \pi_0 \mc{P}_0 + \pi_1 \mc{P}_1$ where $\pi_i$ is the prior probability and $\mc{P}_i$ is the class conditional density of the $i$-th class. The generative model in the main text specifies that $\mc{P}_i = \mc{N}_d\left( (-1)^{i+1}\nu, \Sigma \right)$. For simplicity, we only consider the case where $\pi_0 = \pi_1 = \frac{1}{2}$, but we note that the analysis can be easily extended to unequal priors. Under the 0-1 loss, the risk of an FLD hypothesis $\hat h(x) = \mathrm{1}\{ \hat \omega^\top x > 0 \}$ w.r.t to the target distribution $\mc{P}$ is given by, 
\begin{align*}
    R(\hat h \mid \hat \omega) &= \mathbb{P}_{X \sim \mc{P}}\left[ h(X) \neq Y \mid \hat \omega \right] \\
    &= \frac{1}{2}\mathbb{P}_{X \sim \mc{P}_0}\left[ \hat \omega ^\top X > 0 \right] + \frac{1}{2}\mathbb{P}_{X \sim \mc{P}_1}\left[ \hat \omega ^\top X < 0 \right] \\
    &= \frac{1}{2} - \frac{1}{2}\mathbb{P}_{X \sim \mc{P}_0}\left[ \hat \omega ^\top X < 0 \right] + \frac{1}{2}\mathbb{P}_{X \sim \mc{P}_1}\left[ \hat \omega ^\top X < 0 \right]
\end{align*}
Since $\hat \omega^\top X \sim \mc{N}_1\left( \hat \omega^\top \mathbb{E}[X], \hat \omega^\top \Sigma \; \hat \omega \right)$, we have
\begin{align*}
    R(\hat h \mid \hat \omega) &= \frac{1}{2} - \frac{1}{2}\mathbb{P}\left[ Z < \frac{\hat \omega^\top \nu}{\sqrt{\hat \omega^\top \Sigma \; \hat \omega}} \right] + \frac{1}{2}\mathbb{P}\left[ Z < \frac{-\hat \omega^\top \nu}{\sqrt{\hat \omega^\top \Sigma \; \hat \omega}} \right],
\end{align*}
where $Z$ is a standard normal random variable. Therefore, 
\begin{align*}
    R(\hat h \mid \hat \omega) &= \frac{1}{2} - \frac{1}{2}\Phi \left( \frac{\hat \omega^\top \nu}{\sqrt{\hat \omega^\top \Sigma \; \hat \omega}} \right) + \frac{1}{2}\Phi \left( \frac{-\hat \omega^\top \nu}{\sqrt{\hat \omega^\top \Sigma \; \hat \omega}} \right).
\end{align*}
Using the fact that $\Phi(-x) = 1 - \Phi(x)$, we arrive at the desired expression:
\begin{equation*}
    R(\hat h \mid \hat \omega) = \Phi \left( \frac{-\hat \omega^\top \nu}{\sqrt{\hat \omega^\top \Sigma \; \hat \omega}} \right).
\end{equation*}

\end{document}